\documentclass[iop]{emulateapj}

\usepackage{times}
\usepackage{graphicx}
\usepackage{amsmath}
\usepackage{float}
\usepackage{ifthen}
\usepackage{times}
\usepackage{natbib}
\usepackage{rotating}
\usepackage{units}
\newcommand{\ci}{\,{\rm [C\,{\sc i}]($^3P_1$$\rightarrow^3$$P_0$)}\,}
\newcommand{\cii}{{\rm [C\,{\sc ii}]\,}}%($^3P_1\rightarrow ^3P_0$)}\,}
\newcommand{\coiv}{\,{\rm CO\,{\it J}($4$$\rightarrow$$3$)}\,}

\newcommand{\coi}{\,{\rm CO\,{\it J}($1$$\rightarrow$$0$)}\,}
\newcommand{\coii}{\,{\rm CO\,{\it J}($2$$\rightarrow$$1$)}\,}

\newcommand{\coiii}{\,{\rm CO\,{\it J}($3$$\rightarrow$$2$)}\,}
\newcommand{\hcni}{\,{\rm HCN\,{\it J}($1$$\rightarrow$$0$)}\,}
\newcommand{\hcnii}{\,{\rm HCN\,{\it J}($2$$\rightarrow$$1$)}\,}
\newcommand{\hcniii}{\,{\rm HCN\,{\it J}($3$$\rightarrow$$2$)}\,}

\begin{document}

\shorttitle{Molecular and atomic line surveys of galaxies II}
\shortauthors{Papadopoulos \& Geach}

\title{Molecular and atomic line surveys of galaxies II:\\ Unbiased estimates of their star formation mode}

\author{Padelis \ P.\
Papadopoulos\altaffilmark{1}\ and James~E.~Geach\altaffilmark{2,3}}

\altaffiltext{1}{Max Planck Institute for Radioastronomy, Auf dem H\"ugel 69,
D--53121 Bonn, Germany. padelis@mpifr-bonn.mpg.de}

\altaffiltext{2}{Department of Physics, McGill University,
3600 rue University, Montr\'eal, Qu\'ebec, H3A 2T8, Canada.
jimgeach@physics.mcgill.ca}
\altaffiltext{3}{Banting Fellow}

\begin{abstract}We make use of our `minimal' cold interstellar medium (ISM)
emission line model that predicts the molecular and atomic line emission per
unit dense, star-forming gas mass (Geach \& Papadopoulos\ 2012; Paper\ I) to
examine the utility of key line ratios in surveys of the so-called star
formation `mode' as traced by $\xi_{\rm SF}$=$M_{\rm dense}({\rm H_2})/M_{\rm
total}({\rm H_2})$. We argue that $\xi_{\rm SF}$ and its proxies provide very
sensitive, extinction-free discriminators of rapid starburst/merger-driven
versus secular quiescent/disk-like stellar mass assembly, with the most
promising diagnostic to be applied in the near-future being \coiv/\ci. These
lines are accessible across nearly the full range $0<z<2$ (thus covering the
bulk of galaxy evolution) with the Atacama Large Millimeter Array. In addition
to their diagnostic power, another advantage of this combination is the
similar observed frequencies ($\Delta\nu_0 \approx 30$\,GHz) of the lines,
resulting in nearly spatially--matched beams for a fixed aperture, thus
mitigating the effects of resolution/morphology bias in the interpretation of
galaxy-averaged line ratios. Finally we discuss the capability of deep blind
redshift surveys with the high frequency component of the Square Kilometer
Arrray (SKA) in discovering H$_2$-rich galaxies with very low $\xi_{\rm SF}$
values. These could be the progenitors of starburst galaxies seen prior to the
onset of star formation; such galaxies could be a class of extreme outliers
from local (gas surface density)--(star formation rate) scaling laws, which
would exclude them from current star formation or stellar mass selected
samples. Our conservative model suggests that SKA could detect such systems
residing at $z\sim3$ at a rate of 20--200 per hour.\end{abstract}
\keywords{galaxies: ISM --- galaxies: starburst --- galaxies: evolution ---
cosmology: observations --- ISM: molecules: CO, HCN}

\section{Introduction}

The prevailing `mode' of star formation in the distant Universe, namely
merger-driven starbursts versus isolated star-forming gas-rich disks -- with
their H$_2$ gas as traced by its carbon monoxide (CO) line emission -- is now
a topic of intense interest after the discovery of numerous galaxies at high
redshifts with gas fractions above 50\% that seem to follow the latter mode
(e.g.\ Daddi et al.\ 2010; Tacconi et al.\ 2010). Nevertheless, the indicators
used in such studies, namely the near Galactic value of the $X_{\rm co}$
factor, and low-excitation low-{\it J} CO spectral line energy distributions
(SLEDs) are strongly degenerate when it comes to the star formation mode
(Papadopoulos et al.\ 2012) and thus cannot be used to securely argue for a
bi-modal distribution of star formation modes. Moreover using ratios of single
CO line luminosities (used as measures of total molecular gas mass) to
infrared luminosity (used as star formation rate proxy) as mode indicators is
fraught with uncertainty because the value of the $X_{\rm co}$ remains both
highly uncertain and dependant on galaxy type, molecular gas surface density
and kinematic state, as well as average star formation activity (see Narayanan
et al.\ 2011 for a recent exposition).

In the local Universe the star formation `mode' seems to be very well (i.e.\
uniquely) indicated by the fraction of the total gas reservoir residing in the
dense phase: $\xi_{\rm SF} = M_{\rm dense}({\rm H_2})/M_{\rm total}(\rm H_2)$,
a fact that can be easily exploited in high redshift galaxy surveys. In this
article, we explore practical means of determining the star formation mode of
galaxies, using low-{\it J} CO or \ci and hydrogen cyanide (HCN) line surveys
as the most effective tool for determining $\xi_{\rm SF}$=$M_{\rm dense}({\rm
H}_2)/M_{\rm total}({\rm H}_2)$ as a function of redshift. Selecting galaxies
through $\xi_{\rm SF}$ is independent of the various selection criteria used
in traditional galaxy surveys, and is unaffected by degeneracies of CO SLEDs
in determining $\xi_{\rm SF}$ (Papadopoulos et al.\ 2012).

For the purposes of this work we seek good proxies of the dense gas mass
fraction and investigate their observational prospects in the distant
Universe, especially in survey mode, but will not delve into the details of
how such proxies can be translated into actual $\xi_{\rm SF}$ values.
Multi-{\it J} and multi-species molecular line observations are needed for
this (e.g. Greve et al.\ 2009), which will soon become routine for the Atacama
Large Millimeter Array (ALMA). We note that the prize from actual estimates of
$\xi_{\rm SF}$ values and direct measurements of turbulent line widths in the
molecular gas of distant star-forming galaxies with sensitive imaging
interferometers like ALMA (see Swinbank et al.\ 2011 for an example for a
gas-rich starburst at $z\sim 2.3$) will be to properly investigate proposed
universal turbulence-regulated star formation theory for galaxies (Krumholz \&
McKee\ 2005).

Finally we outline some of the unique discovery possibilities that very deep
low-{\it J} CO ($J_{\rm up}\leq3$) surveys with the Square Kilometer Array
(SKA) can offer when it comes to gas-rich systems with very low $\xi_{\rm SF}$
values that can only be discovered with blind surveys. Such galaxies will have
negligible \hcni and CO $J_{\rm up}\geq 4$ line emission, and would be extreme
outliers of the Schmidt--Kennicutt relation (Kennicutt\ 1998), unselected by
current high-redshift galaxy surveys as these are based, directly or
indirectly, on the star formation rate (SFR, e.g.\ optical/near-IR lines from
H\,{\sc ii} regions, (sub)mm dust thermal continuum, and cm non-thermal
continuum). This population would be valuable for extending our understanding
of the gas properties and early star formation histories of massive galaxies.
Where relevant, we assume a flat cosmological model, with $\Omega_{\rm
m}=0.3$, $\Omega_\Lambda=0.7$ and $H_0=70$\,km\,s$^{-1}$\,Mpc$^{-1}$.

\section{Practical investigations of the star formation mode of galaxies}

\subsection{Theoretical basis}

Paper I presented a model for the abundance of molecular and atomic
line-emitting galaxies seen across cosmic time, using a `minimal' model for
the cold molecular interstellar medium (ISM). We envision that future surveys
will assemble large samples of high-{\it z} systems with bright molecular line
mission, be it in a totally blind sense, or by (sub)mm/cm spectroscopic
follow-up of a SFR (most likely proxied by the submm continuum) or stellar
mass limited sample (achievable with deep optical/near-infrared surveys). A
survey of the star formation mode would be a powerful probe of galaxy
evolution, since it would perform the dual task of measuring the evolution of
the total gas fraction of star-forming galaxies, the nature of the star
formation (burst or secular), and a more accurate prediction of the rate of
consumption of the cold gas reservoirs.

A prominent bi-modality of the $\xi_{\rm SF}$ values in merger-driven versus
isolated disk star formation is actually expected as a general characteristic
of supersonic turbulence because of the much larger ($\sim$5--30$\times$)
turbulent line widths observed in CO line emission in mergers/ultraluminous
infrared galaxies (ULIRGs) than in isolated spiral disks (Downes \& Solomon\
1998). This difference is expected to be {\it a general characteristic of
merger versus isolated disk systems} (Narayanan et al.\ 2011). The resulting
difference in the corresponding 1-dimensional average Mach numbers
$\mathcal{M}_{\rm ULIRGs}\sim (5-30)\times \mathcal{M}_{\rm disks}$ (where
$\mathcal{M}=\sigma _v/c_{\rm s}$) leads to much higher dense gas mass
fractions in the ISM of mergers than in isolated disks given that the
probability distribution function of the molecular gas density in supersonic
turbulence is well approximated by a log-normal distribution with a dispersion
of $\sigma_{\rho} =\surd\left(\ln\left(1+3\mathcal{M}^2/4\right)\right)$
(Padoan \& Nordlund 2002).

The very enhanced dense gas mass fractions molecular gas mass fraction
expected to be in structures with overdensities $x\geq x_{0}$ (where $x=
n/\langle n \rangle$, and $n$ is the gas density) is then given by

\begin{equation}
\xi_{\rm SF}=\frac{M(x\geq x_{0})}{M_{\rm total}}=
\frac{1}{2}\left[1+{\rm erf}\left(\frac{-2\ln(x_{0})+\sigma^2 _{\rho}}{2^{3/2} \sigma_{\rho}}\right)\right].
\end{equation}

The typically large velocity dispersions of
$\sigma_{v}$$\sim$(30--140)\,km\,s$^{-1}$ in the molecular gas disks of the
merger systems found in ULIRGs (and recently even in a molecular gas disk
fueling star formation in an spectacular starburst at $z=2.3$, see Swinbank et
al.\ 2011), with corresponding large $\mathcal{M}$ will dramatically extend
the gas density probability distribution function towards high values where
much of the molecular gas mass will lie, quite unlike the less turbulent
molecular gas in isolated disks (see Figure\ 1). Thus for giant molecular
clouds (GMCs) in typical spiral disks only $\sim$3\% of their mass is expected
to be at overdensities $x>500$ (5$\times$10$^4$\,cm$^{-3}$ for a typical GMC
with $\langle n\rangle=100$\,cm$^{-3}$) while for the highly turbulent gas in
mergers this is $\sim$45\%, and can be even higher (e.g. Greve et al.\ 2009).

\begin{figure}[t]
\centerline{\includegraphics[width=0.5\textwidth,angle=-90]{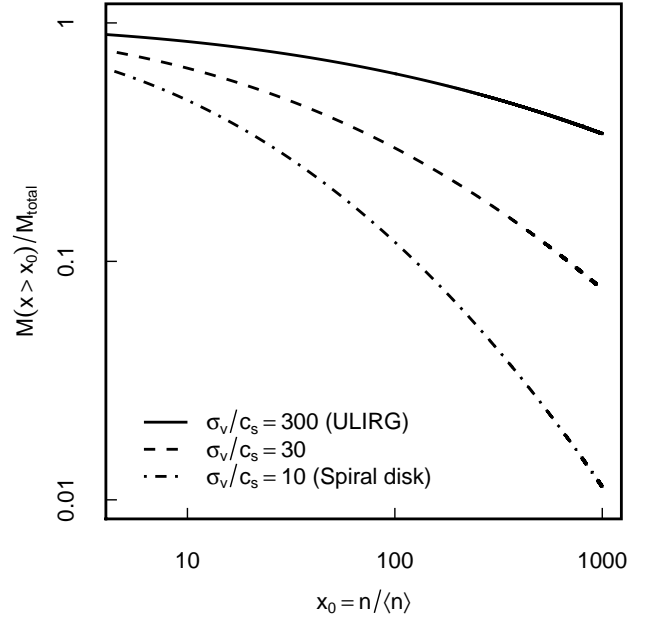}}
\caption{The gas mass fraction expected at overdensities of $x\geq x_{0}$
(where $x=n/\langle n \rangle$, see Equation 1) in Giant Molecular Clouds
(GMCs) for a given 1-dimensional average Mach number
$\mathcal{M}=\sigma_v/c_s$. The large boost of $M(x\geq x_{0})/M_{\rm total}$
that occurs in GMCs resident in merger/ULIRGs ($\mathcal{M}\approx(30-300)$) with respect to those in spiral disks ($\mathcal{M}\approx10$) is evident.} \end{figure}

\subsection{Practical tracers of the star formation mode}

In the following we consider various combinations of molecular and atomic
lines as proxies of $\xi_{\rm SF}$. A summary from available line data for
mostly local galaxies can be found in Figure\ 2 where we plot normalized line
ratios for a sample of observations of (i) quiescent star forming disks and
dark clouds and (ii) galactic nuclei, mergers and starburst galaxies, compiled
from the literature (Gao \& Solomon\ 2004, Israel et al.\ 1998, 2001, 2003,
Barvainis et al.\ 1997, White et al.\ 1994, Petitpas \& Wilson\ 1998). We
define the normalized ratios as $r_{\rm a/b}/\langle r_{\rm a/b} \rangle$,
where ${\rm a}$ and ${\rm b}$ correspond to (for example) the line brightness
luminosities of the dense and total gas tracers \hcni and \coi with $\langle
r_{\rm a/b} \rangle$ representing the average over the entire sample.

It is clear from Figure\ 2 that carefully-picked line diagnostics can
sensitively distinguish between `disk' and `burst' mode, with $r_{\rm
a/b}/\langle r_{\rm a/b} \rangle \approx 1$ dividing the two classes.
Individual $\xi _{\rm SF}$ tracers (e.g.\ \hcni and \coi line ratios) can be
even more sensitive than what the distribution in Figure\ 2 indicates, with
factors of $\sim$10 difference between `disk' and `burst' mode (see \S2.2.1).
Moreover, the sample presented in Figure\ 2 is dominated by local systems
while high redshift observations present more of a challenge as global line
measurements may represent wide spatial averages (smearing out the line ratio
for composite systems), but also because the observed frequencies of lines
available in the local Universe will be redshifted out of the current range of
sensitive facilities for certain important cosmological spans of time.

We now will now discuss several of these tracers in more detail, highlighting
their relative merits with a view to identifying diagnostic tools that will be
suitable for practical surveys of the star formation mode of galaxies over
large spans of cosmic time.

\subsubsection{\hcni/\coi, \hcnii/\coii}

This is the cleanest $\xi_{\rm SF}$ indicator available, with local studies
showing a clear bi-modality for the ISM in merger-driven starbursts versus
isolated disks with lower SFRs in the local Universe (Gao \& Solomon\ 2004),
with HCN tracing the dense, star-forming gas mass, and low-{\it J} CO tracking
the total gas reservoir. Arguably this line combination is the most direct
tracer of $\xi_{\rm SF}$; extensive observations of local galaxies show this
line ratio to be an excellent way to discriminate between different modes,
with high values of $r^{(10)}_{\rm HCN/CO}\sim 0.2$--$0.3$ found exclusively
in compact, merger-driven, extreme starbursts while $r^{(10)}_{\rm HCN/CO}\sim
0.01$--$0.03$ for isolated spiral disks with extended star formation (e.g.\
Solomon et al.\ 1992; Gao\ \& Solomon\ 2004).

Observing \hcni ($\nu_0=88.632$\,GHz) and \coi ($\nu_0=115.271$\,GHz) has the
disadvantage that, although this line combination will be accessible with ALMA
for local systems (important for an even more robust calibration in terms of
$\xi_{\rm SF}$), by $z\sim1$ \coi is redshifted out of the ALMA bands and
remains inaccessible until $z\sim1.5$ in the cm bands (and accessible with
other cm-wave facilities). In addition, due to the broad telluric absorption
feature at $\nu\sim60$\,GHz, there will be a broad \hcni and \coi `desert' in
the $z<1$ interval, limiting the use of this tracer for intermediate redshift
work.

Due to the aforementioned shortcomings, this particular indicator will only be
effective for probing very local galaxies rather than those at cosmologically
significant redshifts of $z>2$, and thus cannot satisfactorily measure
$\xi_{\rm SF}$ over the peak of the global star formation rate density in a
well-sampled, tomographic manner. However, given the small range of both CO
and HCN $r_{21}\sim 0.6$--$1$, the $J=2\rightarrow 1$ transition of these two
molecules can also be used to the same effect. This allows ALMA to survey
$r_{\rm HCN/CO}$ up to $z\la 1$ as the rest frequencies of \coii and \hcnii
are at 230.538\,GHz and 177.264\,GHz respectively. \hcnii can be accessed by
ALMA out to $z\sim1$, and \hcniii out to $z\sim2$. Given the order of
magnitude differences expected for the $r_{\rm HCN/CO}$-type of ratios between
mergers and isolated disks it is very likely that even \hcniii and ratios such
as \hcniii/\coi or \hcniii/\coii will remain practical proxies of $\xi_{\rm
SF}$ out to $z\lesssim2$ using ALMA. Moreover, given the tremendous
sensitivity boost at (sub)mm wavelengths that ALMA represents, the necessary
calibration of such `hybrid' HCN/CO ratios in terms of $\xi_{\rm SF}$ will be
trivial for large numbers of star-forming galaxies in the local Universe.

\subsubsection{\rm \hcni/\ci, \hcnii/\ci} 

This is a potentially powerful tracer of $\xi_{\rm SF}$, with -- as mentioned
above -- the \hcni and \hcnii line luminosities proportional to the dense
star-forming gas mass, and \ci scaling with total molecular gas mass, but with
many advantages over low-{\it J} CO lines (see Paper\ I). The higher rest
frequency of \ci ($\nu_0=492$\,GHz), compared to \hcni means that the
corresponding \ci observations would be conducted with ALMA Bands 6 and lower.

Local data support the notion of the ratio of \hcni and \ci lines as a
$\xi_{\rm SF}$ proxy, with $r_{\rm HCN/[CI]}$$\sim$0.55--2.2 observed in
ULIRGs (\ci data presented in Papadopoulos\ \&\ Greve\ 2004; Greve\ et al.\
2009, and \hcni data from Gao\ \&\ Solomon\ 2004). In comparison, for
disk-dominated quiescent galaxies, $r_{\rm HCN/[CI]}\sim 0.06-0.10$, although
it can rise up to $\sim$0.5 for vigorously star-forming disks like that of
NGC\,1068 (using \hcni data from Gao \& Solomon [2004] and \ci data from
Israel et al.\ [2009]). Further benchmarking work on this ratio as a practical
mode indicator must be done locally (now possible from high altitude dry sites
like the Chajnantor plateau) before applying it in the distant Universe. A
larger body of multi-{\it J} HCN line data along with \ci measurements and
models of the $\rm [C/H_2]$ abundance\footnote{It must be stressed that
abundance uncertainties affect {\it all} H$_2$ mass measurements utilizing
optically thin lines (e.g.\ $^{13}$CO). Among those atomic carbon is the
species with the simplest, mostly cosmic ray controlled chemistry.} are
necessary to both benchmark $r_{\rm HCN/[CI]}$-type ratios as $\xi_{\rm SF}$
indicators as well as help disentangle the effects of high SFRs (and thus
[C/H$_2$]-boosting cosmic rays) on this valuable mode indicator (see also
\S2.2.4).

\subsubsection{\rm $^{12}$CO/$^{13}$CO ratios for $J_{\rm up}\leq2$}

The $R_{\rm 12/13}(1\rightarrow0,\,2\rightarrow1)={\rm^{13}CO/^{12}CO}$ ratios
are not proxies of $\xi_{\rm SF}$ but rather of the turbulent velocity fields
of giant molecular clouds in star-forming galaxies. Numerous studies have
shown that $R_{\rm 12/13}(1\rightarrow0, \,2\rightarrow1)\sim5-15$ in disks,
with low values found in Milky Way type systems and larger ones in vigorously
star-forming disks (e.g.\ Casoli et al.\ 1992; Aalto et al.\ 1995;
Papadopoulos \& Seaquist\ 1999). Ratios $R_{\rm
12/13}(1\rightarrow0,\,2\rightarrow1)\geq 20$ are found {\it exclusively} in
merger systems and are thought to be the result of the highly turbulent ISM
and the enhanced gas temperatures produced by the dissipation of their
supersonic turbulence (Aalto et al.\ 1995; Greve et al.\ 2009 and references
therein).

The advantage of using isotopologue line ratios as proxies of merger versus
isolated-disk star formation modes, even if such ratios are not proxies of
$\xi_{\rm SF}$, is obvious: the convenience yielded by their frequency
proximity. Wide bandwidth receivers allow the possibility of deep surveys with
simultaneous CO and $^{13}$CO line observations. Objects with
$R_{12/13}(1\rightarrow0,\,2\rightarrow1)\geq 20$ would be considered as
harbouring merger-driven star formation, while all the rest would be in the
disk mode. This method can safely be used up to $J_{\rm up}=2$, partly because
of lack of local $^{13}$\coiii or higher-{\it J} line data (the $^{13}$\coiii
line lies precariously close to an atmospheric absorption feature) but also
because higher-{\it J} CO lines will progressively cease to be good tracers of
the {\it average} velocity fields of molecular clouds and begin tracing their
self-gravitating, dense regions (Ossenkopf\ 2002). Nevertheless recent
measurements of a large $R_{12/13}(J=3\rightarrow2)$ in a distant
ULIRG-hosting QSO (Henkel et al.\ 2010) indicate that such a diagnostic may
remain a good star formation mode indicator up to $J_{\rm up}=3$.

Finally we note that a competing explanation for the very high
$R_{12/13}(1\rightarrow0,\, 2\rightarrow1)$ ratios observed in all ULIRGs
exists, attributing them to much higher $\rm [^{12}C/^{13}C]$ abundances in
such systems (Casoli et al.\ 1992; Henkel \& Mauersberger\ 1993 and references
therein), possibly caused by a different stellar initial mass function in
merger/starburst systems. If this is true it would make $R_{12/13}$ a less
`clean' star formation mode indicator unless a strong link can be proven
between such isotopic enhancements and merger-driven star-forming systems.

\begin{figure}[t]
\centerline{\includegraphics[width=0.5\textwidth,angle=-90]{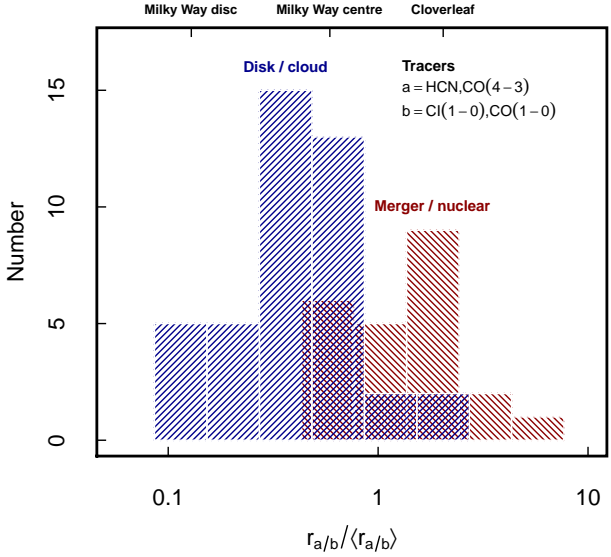}}
\caption{Comparison of the $\xi_{\rm SF}$ proxies $r_{\rm HCN/CO}$, $r_{\rm
HCN/[CI]}$ and $r_{\rm CO(4-3)/[CI]}$ compiled for galaxy disks and clouds
(representing our `disk mode') and mergers and nuclear regions (representing
our `burst mode'); data taken from Gao \& Solomon\ (2004), Israel et al.\
(1998, 2001, 2003), Barvainis et al.\ (1997), White et al.\ (1994), Petitpas
\& Wilson\ (1998). To present the tracers on the same scale, we have
normalized each proxy to the average value, $r_{\rm a/b}/\langle r_{\rm a/b}
\rangle$, where ${\rm a}$ and ${\rm b}$ correspond to (e.g.) \hcni and \coi.
Although there is some overlap, due to the $r_{\rm CO(4-3)/[CI]}$ and $r_{\rm
HCN/[CI]}$ ratios (see discussion in \S2.2.2 and \S2.2.4) there is a clear
distinction between the two classes of systems irrespective of the particular
indicator used, with $r_{\rm a/b}/\langle r_{\rm a/b} \rangle \lesssim1$ for
quiescent disks and $r_{\rm a/b}/\langle r_{\rm a/b} \rangle\gtrsim1$ for
merger/nuclear systems.} \label{fig:sled} \end{figure}

\subsubsection{\coiv/\ci} 

The $r_{\rm CO(4-3)/[CI]}$ ratio has the key advantage that the lines used
have similar rest frequencies, ($\Delta \nu_0 \approx 30$\,GHz, and thus
similar beams for a given aperture), while both remain accessible to ALMA
across a large span of cosmic time. It is assumed that \coiv traces mostly
dense star-forming gas, an expectation borne out both by observations (e.g.\
Petitpas \& Wilson 1988; Nieten et al.\ 1999; Papadopoulos et al.\ 2012) as
well as models of emergent CO line emission from turbulent molecular clouds
(Ossenkopf\ 2002). Finally, both lines benefit from a positive {\it
k}-correction with respect to lower-{\it J} lines making them suitable for
(H$_2$ mass)-sensitive high-{\it z} work.

Unfortunately, their relatively high frequencies and the lack of multi-beam
receivers at such frequencies has prevented the assembly of a large body of
\coiv and \ci line data in the extragalactic domain. Nevertheless, enough data
is assembled to examine if a significant bi-modality of $r_{\rm CO(4-3)/[CI]}$
indeed exists for starburst environments dominated by large dense gas
reservoirs (typical of merger-driven systems) versus low-SFR environments in
disks. In this comparison we also consider \coiv and \ci data obtained for
galactic nuclei (e.g.\ NGC\ 253) including that of the Galactic Center as
representative of the {\it average} molecular ISM properties of the
merger-driven starbursts in ULIRGs (the high turbulence yielding also very
high $M_{\rm dense}/M_{\rm total}$ values in galactic centers). For such
environments we obtain $\langle r_{\rm CO(4-3)/[CI]}\rangle =4.55\pm 1.5$
(data from Israel et al.\ 1995,\ 1998,\ 2001,\ 2003; Petitpas \& Wilson\ 1998;
Barvainis et al.\ 1997; Papadopoulos \& Greve\ 2004; Papadopoulos et al.\
2004) where the dispersion represents that of the measured values. These
include the Galactic Center, the nuclei of various nearby galaxies (NGC\, 253,
Maffei\,2, M\,83), a distant strongly-lensed QSO (The Cloverleaf at $z=2.56$)
and a local ULIRG/QSO (Mrk\,231). For disk-dominated star-forming
environments, the available literature data yield $\langle r_{\rm
CO(4-3)/[CI]}\rangle\sim0.45$--$1.3$ with the lowest value ($\sim$0.43) found
for the outer Galaxy and a quiescent cloud in M\,31 ($\la$0.45). Somewhat
higher values are obtained for the inner Galaxy ($\sim$0.83) and star-forming
spirals such as NGC\,4826 and NGC\,3079 ($\sim$1.3, and see Figure\ 2).

Thus it is seems that the $r_{\rm CO(4\rightarrow3)/[CI]}$ ratio {\it can} be
a practical star formation mode indicator in the ALMA regime, with differences
up to factors of $\sim$10 between mergers and disk-dominated systems. We note
however that the intense cosmic ray fluxes expected in starburst environments
can significantly boost the [C{\sc i}/H$_2$] abundance ratio (e.g.\
Papadopoulos et al.\ 2004 and references therein), reducing the $r_{\rm
CO(4\rightarrow3)/[CI]}$ ratio in extreme starbursts. This may be happening in
the inner 500\,pc of M\,82 where a rather low $r_{\rm
CO(4\rightarrow3)/[CI]}=1.5$ is measured (White et al.\ 1994). Chemical models
of [C{\sc i}/H$_2$] as a function of average cosmic ray energy densities
($\propto$SFR density) are necessary to account for such effects as well as
reducing the [C{\sc i}/H$_2$] abundance uncertainties (also needed for $M_{\rm
total}$(H$_2$) estimates when using the the \ci line luminosity). We expect
that even crude models of a SFR-dependant (i.e.\ cosmic ray-dependant) [C{\sc
i}/H$_2$] average abundance will yield $r_{\rm
CO(4\rightarrow3)/[CI]}$\,(cosmic ray$=$Galactic) ratios (i.e.\ referenced to
a common Galactic cosmic ray energy density) that will be even more sensitive
to $\xi_{\rm SF}$ (and thus the star formation mode) by much reducing the
little overlap that exists for this currently cosmic ray-uncorrected star
formation mode proxy (and $r_{\rm HCN/[CI]}$) in Figure\ 2.

\subsection{\coiv and \ci line flux predictions and the design of a practical star formation mode survey}

In light of the distinct advantages that this diagnostic offers for high-{\it
z} studies, in Figure\ 3 we plot the fluxes of the \ci and \coiv lines for a
$L_{\rm IR}=5\times10^{11}L_\odot$ galaxy at $0<z<2$ undergoing a burst
($\xi_{\rm SF}=0.6$) and disk-mode star formation ($\xi_{\rm SF}=0.04$). To
calculate the line fluxes, we use our minimalist ISM model presented in Paper
I, assuming the `supervirial' and `virial' versions of the model for the burst
and disk mode respectively (see \S3 of Paper I for further details). The \ci
line flux is estimated from the total gas mass following equation 11 of
Papadopoulos, Thi \& Viti\ (2004), adopting an excitation value of
$Q_{10}=0.35$ (where $Q_{10}=N_{\rm 10}/N_{\rm [CI]}$ is the ratio of the
column density of gas in level $^3P_1$ to the ground state, appropriate for
the lower temperature and excitation conditions when averaged over a global
gas reservoir. The line ratios predicted from our model are $r_{\rm
CO(4\rightarrow3)/CI} = 0.3$ for $\xi_{\rm SF}=0.04$ and $r_{\rm
CO(4\rightarrow3)/CI} = 2.1$ for $\xi_{\rm SF}=0.6$; in agreement with the
normalized values presented in Figure\ 2.

We compare the predicted line fluxes to the 5$\sigma$ sensitivity (for
channels of 300\,km\,s$^{-1}$) of full-power ALMA (assumed to be 50 antennas)
in integrations of 1, 10 and 24\,hours. This highlights the suitability of
this line combination as a tool for surveys of the star formation mode, with
both \coiv and \ci lines detectable in realistic times for relatively normal
galaxies over the full span of cosmological time pertinent to galaxy evolution
in both disk and burst modes. The similarity of the line frequencies are key,
because they provide nearly spatially matched beams (for a given dish, and in
the case of interferometers, array configurations\footnote{In Paper I and
here, we have assumed the most compact array configuration will be used in
survey mode.}), and this can be critical in mitigating the effect of any
structural bias in the interpretation of such line ratios due to the `mixing'
of line emission from systems with different star formation modes that may
exist in relative close proximity in merger configurations (e.g.\ Ivison et
al.\ 2010). Moreover, in the cases of well-resolved individual systems the
$r_{\rm CO(4\rightarrow3)/CI}$ ratio can give a quick approximate `reading' of
the $\Sigma_{\rm dense}/\Sigma ({\rm H}_2)$ gas surface density ratio as a
function of position within the galaxy. This level of detail, while certainly
of interest to other types of studies concerning, for example, the detailed
gas dynamics and its inward transport in galactic disks, will still be
valuable for uncovering `milder' merger systems where the original disk is not
destroyed but its $M_{\rm dense}({\rm H}_2)/M_{\rm total}({\rm H}_2)$ is
nevertheless enhanced by minor merger events.

\begin{figure}[t]
\centerline{\includegraphics[width=0.5\textwidth,angle=-90]{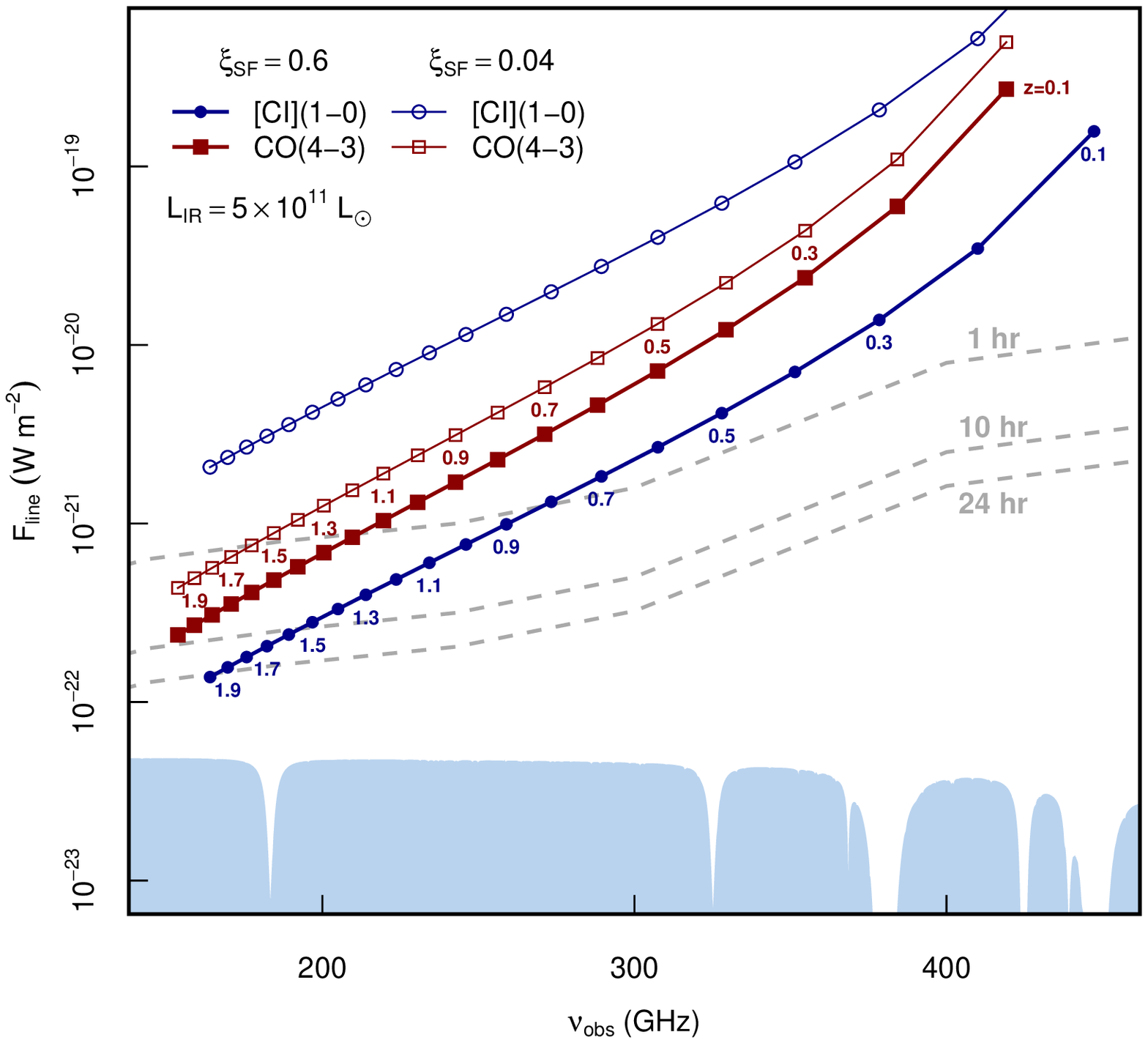}}
\caption{Predicted \coiv and \ci line fluxes for a galaxy with $L_{\rm
IR}=5\times10^{11}L_\odot$ in the burst mode ($\xi_{\rm SF}=0.6$) and disk
mode ($\xi_{\rm SF}=0.04$) seen at $0<z<2$. Our line flux estimates are based
on our `minimal' ISM model presented in Paper I. The benefit of using this
line combination as a proxy for $\xi_{\rm SF}$ is in the similarity of line
frequencies, and so spatially matched beams for a given telescope, and the
near-continuous accessibility of the lines over cosmological distances from
the ground (the blue shaded region shows the atmospheric transmission -- on an
arbitrary vertical scale -- for a column of 1\,mm of precipitable water vapour
at Chajnantor, indicating the available windows for observations). Here we
show the 5$\sigma$ sensitivity of full power ALMA (50 antennas) achieved in 1,
10 and 24\,hours, demonstrating the feasibility of performing a star formation
mode survey for a luminosity-selected sample of galaxies over the majority of
cosmic history, covering the peak epoch of star formation to the present day.
} \label{fig:sled} \end{figure}

\section{Searching for H$_2$-rich galaxies with ultra-low star formation rates
in the distant Universe}

One of the most exciting aspects of deep blind molecular line surveys is their
potential to discover gas-rich systems with very low levels of star formation;
i.e.\ very low $\xi_{\rm SF}$. This could be a new class of objects that may
be currently omitted by even the deepest galaxy surveys as these depend always
on the SFR either directly (e.g.\ optical/near-infrared lines from H\,{\sc ii}
regions) or indirectly (non-thermal cm continuum, (sub)mm dust continuum). The
SKA and its pathfinders can search for the H\,{\sc i} phase of such systems
while deep low-{\it J} CO and \ci line observations (the latter with ALMA) can
reveal their molecular gas reservoir. Of course the existence of the latter
and its detectability via CO and \ci lines assumes some significant metal
enrichment of its ISM (so H$_2$ can form within reasonable time and CO, \ci
can be there to trace it) and thus some initial star formation episode. This
need not contradict very low SFRs in the subsequent evolution of such systems,
as a combination of structural properties (e.g.\ large thin disks), star
formation feedback, and very different average ISM properties (very low
density molecular clouds with little molecular cooling) may act to suppress
SFRs over long periods of time.

This has already been shown possible, albeit for short periods of time, for
gas-rich and metal-poor disks with otherwise typical structural
characteristics (Papadopoulos \& Pelupessy\ 2010). During such periods these
systems can undergo `bursts' of very inefficient SF (as well as `bursts' of
very efficient star formation) that would place these systems as extreme
deviants of the local Schmidt-Kennicutt (S--K) SFR/gas density scalings.
Regardless whether H$_2$-rich galaxies with ultra-low SFRs are simply a) short
(S--K)-deviant periods over much longer lifetimes of otherwise normal SF
galaxies (e.g.\ `pre-starbursts': gas-rich systems seen just prior to a burst
of intense star formation), or b) a completely new class of objects widely
spread across cosmic epoch, deep low-{\it J} CO and \ci blind line surveys are
the way to find them. The \ci line is particularly promising because of its
large positive {\it k}-correction (with respect to low-{\it J} CO lines) yet
easy excitation, and its potential to trace the metal-poor ISM with much of CO
dissociated by far-UV radiation (or simply never formed in sufficient
abundances) leaving \ci and \cii as the dominant form of carbon. On the other
hand, the SKA will have the sensitivity and field of view making it favourable
for searching for low-{\it J} CO at high-redshift compared to \ci in the ALMA
bands. We note that the \cii line can trace such ISM much more sensitively
than \ci but unfortunately it becomes accessible to sensitive ALMA bands only
for $z\ga 2$ (Figure\ 2 in Paper\ I), and it cannot be easily ´´translated´´
to total molecular gas mass like the latter.

Here we consider a na\"ive model, where it is assumed that every star-forming
galaxy goes through a brief pre-starburst phase prior to the onset of star
formation, such that the abundance of sources is simply proportional to number
of galaxies contributing to the infrared luminosity density used in our
abundance model (Paper\ I): $\Phi_{\rm PSB}=\lambda \Phi_{\rm IR}(z,L)$. We
assume a pre-starburst `duty cycle' of $\lambda=0.1-1$\% and `low' mode of
$\xi_{\rm SF}=0.01$, such that the majority of gas is in the cold, quiescent
phase. Clearly the SKA with its sensitivity, wide field of view and access to
low-{\it J} CO lines at $z>3$ is the most suitable facility for a search of
pre-starburst galaxies. We present the cumulative number counts of \coi
emitters at $\nu_{\rm obs}=30$\,GHz (assuming an instantaneous bandwidth of
4\,GHz) and $\lambda=0.1-1$\% in Figure\ 4. Operating at the `optimal' flux
limit that maximizes the detection rate for a fixed observing time (i.e.\
where the integral counts are proportional to $F^{-2}$, see Blain et al.\
2000; Carilli \& Blain\ 2002), we predict that a blind survey with SKA could
detect these S--K outliers at $z\approx3$ at rate of 20--200 galaxies per
hour\footnote{assuming the latest design specifications for a high-frequency
SKA: skatelescope.org, S. Rawlings, private communication, 2011}. This model
is certainly too simplistic; cosmological simulations that include a
prescription for the chemistry and the evolution of the cold molecular gas
might be better placed to make predictions for the abundance and detectability
of S--K outliers with ultra-low SFRs at high-$z$ (Obreschkow et al.\ 2009;
Obreschkow, Heywood \& Rawlings, 2011; Lagos et al.\ 2011, 2012; J.\ E.\ Geach
et al.\ 2012 in preparation).

Finally we note that imaging of dense gas tracers such as HCN and carbon
monosulfide (CS) line emission along with indicators of total molecular gas
distribution such as \coi, \coii and \ci lines is of paramount importance for
extracting information on important evolutionary parameters of galaxies such
as $ M_{\rm dyn}/M_{\rm total}({\rm H}_2)$ (with more gas-rich systems
expected at high redshifts irrespective of the star formation mode) and gas
consumption timescales $\tau _{\rm SF}=M_{\rm dense}({\rm H}_2)/{\rm SFR}$.
Indeed it is only for the dense gas phase that the latter have the intended
meaning as the duration of the observed star formation event, though in the
literature the longer timescales of $\tau \sim M_{\rm total}({\rm H}_2)/{\rm
SFR}$ are often used, a rather poor choice as not all of the molecular gas is
relevant to star formation, only the dense phase.

\begin{figure}[t]
\centerline{\includegraphics[width=0.5\textwidth,angle=-90]{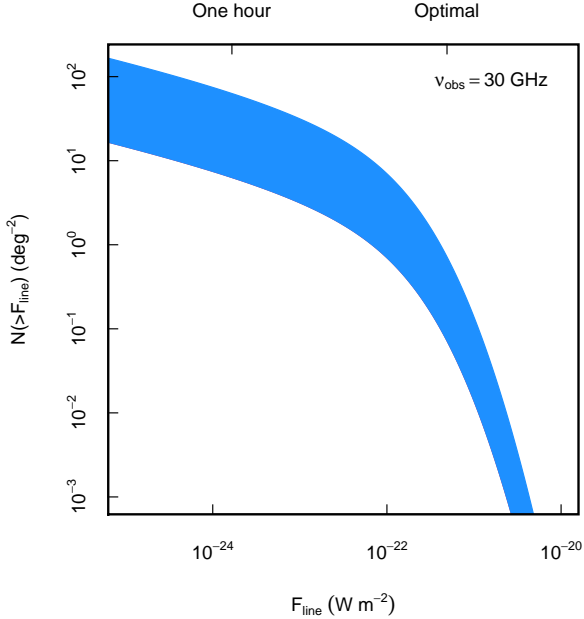}}
\caption{Predicted abundance of \coi emitters at $\nu_{\rm obs}=30$\,GHz
representing `pre-starburst' galaxies, i.e.\ gas rich galaxies with $\xi_{\rm
SF}=0.01$ (i.e.\ very low values) at $z\approx3$. The abundance estimates are
calculated using the the model presented in Paper I. Our model ties the number
density of CO emitters to the evolution of the infrared luminosity density
(B\'ethermin et al.\ 2011), relating the luminosity of the principle molecular
and atomic lines using our minimal ISM model. To predict the abundance of
these pre-starbursts, we assume a na\"ive model where the space density of
pre-starbursts is related to the infrared luminosity density like $\Phi_{\rm
PSB}=\lambda \Phi_{\rm IR}(z,L)$. In this plot we show the \coi surface
density for a range $\lambda=0.1$--$1$\%. Assuming the latest design
specifications for the SKA high frequency component, operating at the
`optimal' flux limit -- where the counts become proportional to $F^{-2}$ -- we
expect the blind discovery of 20--200 of these S--K outliers per hour. }
\end{figure}

\section{Conclusions and summary}

In this work we use our minimal molecular and atomic line emission model
presented in Paper\,I (where it was used to assess the practicality of blind
galaxy surveys using ISM molecular and atomic lines), to discuss practical
surveys of the star formation `mode' of galaxies, i.e.\ isolated disks versus
mergers, using the mass fraction $\xi_{\rm SF}$ of the dense star-forming gas
phase versus the total molecular gas mass residing in star-forming galaxies.
We argue that theoretical expectations as well as local molecular line studies
point to $\xi_{\rm SF}$ as a very sensitive, extinction-free, measure of the
star formation mode, independent of any galaxy morphological criteria.
Moreover $\xi_{\rm SF}$ does not suffer from well-known degeneracies of
low-{\it J} CO lines and the values of the so-called $X_{\rm CO}$ factor which
make them poor star formation mode indicators of galaxies. Among the various
proxies of $\xi_{\rm SF}$ we discuss, including HCN/CO ($J_{\rm up}\leq2$),
\hcni/\ci, ratios, we find that the \coiv/\coi\ ratio currently offers the
most immediately practical path to star formation mode surveys using ALMA,
with the similarity of the lines' frequencies offering observing efficiency,
and the nearly matched beam size for a given telescope (the most suitable
being ALMA for galaxy evolution studies) being an important advantage.
Sensitive $^{12}$CO/$^{13}$CO for $J_{\rm up}\leq2$ line ratio surveys are
another promising route maintaining the latter advantage, though such ratios
are not direct proxies of $\xi_{\rm SF}$ rather than of the average turbulence
levels in the ISM environments of disks (with low $^{12}$CO/$^{13}$CO ratios)
and mergers (with high $^{12}$CO/$^{13}$CO ratios).

We note that for uniquely parametrizing the $\xi_{\rm SF}$ in gas-rich
star-forming systems in the distant Universe, the specific high-density gas
tracers used are not important. As sensitivities dramatically improve in the
era of ALMA, other tracers will become available; for example, CS lines will
also become suitable, with the same merger-versus-disk bi-modality expected
for the corresponding $r_{\rm CS/CO}$ and $r_{\rm CS/[CI]}$ ratios. Of course
before any foray into the high redshift Universe any proxy of $\xi_{\rm SF}$
must be calibrated locally with large samples of star-forming galaxies, a task
easily performed with the vastly improved sensitivities of ALMA for all
typical dense (and total) molecular gas mass tracers.

Finally we comment on the capabilities of upcoming observing facilities in
searching for H$_2$-rich galaxies with ultra-low star formation rates in the
distant Universe. These would be extreme outliers of the local (SFR
density)-(gas mass) Schmidt-Kennicutt (S--K) relations, and deep low-{\it J}
CO and \ci line observations with the SKA and ALMA are the only way to
discover them. The \cii line can also be used to trace such systems very
effectively with ALMA, especially if they are metal-poor, but only at very
high redshifts where the line is redshifted into the sensitive ALMA bands
(Paper\ I), while its luminosity cannot be straightforwardly assigned to total
molecular gas mass. Future cosmological simulations of galaxy evolution that
include their molecular gas phase can inform on such possibilities, but it
must be emphasized that such a putative population of (S--K) relation outlier
gas-rich galaxies with ultra-low SFRs represents pure discovery space for deep
molecular and atomic ISM line surveys using ALMA, SKA and its precursors, and
thus should certainly be undertaken.

 P.P.P.\ would like to thank the Director of the Argelander Institute of
Astronomy Frank Bertoldi, the Rectorate of the University of Bonn, and the
Dean U.-G. Meissner, for their `Hausverbot' initiative that was a catalyst for
finishing this work ahead of schedule. J.E.G.\ is supported by a Banting
Postdoctoral Fellowship administered by the Natural Sciences and Engineering
Research Council of Canada.

\label{lastpage}

\end{document}